\documentstyle[aps,prb,graphics,multicol,epsfig]{revtex}

\def\bs#1{\bbox{#1}}
\def\ve#1{\bs{\mathbf{#1}}}
\newcommand{\be}{\begin{eqnarray}}
\newcommand{\ee}{\end{eqnarray}}

\begin{document}

\draft \title{Electromagnetic force on a metallic particle in presence
of a dielectric surface} \author{P. C. Chaumet and
M. Nieto-Vesperinas}

\address{Instituto de Ciencia de Materiales de Madrid, Consejo
Superior de investigaciones Cient\'{\i}ficas, Campus de Cantoblanco
Madrid 28049, Spain}

\maketitle

\begin{abstract}   

By using a method, previously established to calculate electromagnetic
fields, we compute the force of light upon a metallic particle. This
procedure is based on both Maxwell's Stress Tensor and the Couple
Dipole Method. With these tools, we study the force when the particle
is over a flat dielectric surface. The multiple interaction of light
between the particle and the surface is fully taken into account. The
wave illuminating the particle is either evanescent or propagating
depending an whether or not total internal reflection takes place. We
analyze the behaviour of this force on either a small or a large
particle in terms of the wavelength. A remarkable result obtained for
evanescent field illumination, is that the force on a small silver
particle can be either attractive or repulsive depending on the
wavelength. This behaviour also varies as the particle becomes larger.

\end{abstract}

                                                              
\begin{multicols}{2}

\section{Introduction}

Since the first demonstration of particle manipulation by the action
of optical forces,~\cite{ashkin69,ashkin70} optical
tweezers~\cite{ashkin86} and other configurations of light beams have
been established to hold suspended particles like molecules
,~\cite{ashkin87} or more recently, dielectric
spheres.~\cite{collins,gauthier,taguchi} Also, the possibilities of
creating microstructures by optical binding and resonance effects have
been discussed~\cite{burns,anto1,anto2,bayer,novotny} as well as the
control of particles by evanescent waves.~\cite{kawata,sugiura} Only a
few works exist on the interpretation, prediction and control of the
optical force acting on a small particle on a plane surface. To our
knowledge, the only theoretical works dealing with this subject are
those of Refs.~[\ref{almaas},\ref{lester},\ref{prbchaumetnieto}]. In
Ref.~[\ref{almaas}] no multiple interaction of the light between the
particle and the dielectric surface is considered. On the other hand,
Ref.~[\ref{lester}] deals with a 2-D situation. Only recently in
Ref.[\ref{prbchaumetnieto}] the full 3-D case with multiple scattering
was addressed for dielectric particles.

This work, extends the study of Ref.[\ref{prbchaumetnieto}] to
metallic particles and, as such, this is the first theoretical study
of light action on a metallic particle. We shall therefore present a
rigorous procedure to evaluate the electromagnetic force in three
dimensions. Further, we shall analyze how this force depends on the
wavelength, distance between the particle and the surface, angle of
incidence (whether the excitation is a plane propagating or an
evanescent wave), and on the excitation of plasmons on the sphere. We
shall make use of the couple dipole method previously employed, and
whose validity was analyzed in detail in Ref.[\ref{prbchaumetnieto}].

In section~\ref{forceg} we introduce a brief outline on the method
used to compute the optical force on a particle. We also write its
expression from the dipole approximation for a metallic sphere in
presence of a surface. Then, in Section~\ref{petite} we present the
results and discussion obtained in the limit of a small sphere, and in
Section~\ref{grande} we analyze the case of larger spheres compared to
the wavelength.

\section{Computation of the optical forces}\label{forceg}

The Coupled Dipole Method (CDM) was introduced by Purcell and
Pennypacker in 1973.\cite{purcell} In this paper we use this procedure
together with Maxwell's Stress Tensor (MST)~\cite{stratton} in order
to compute the optical forces on a metallic object in the presence of
a surface. Since we developed this method in a previous
paper,~\cite{prbchaumetnieto} we shall now outline only its main
features. It should be remarked that all calculations next will be
written in CGS units for an object in vacuum.

The system under study is a sphere, represented by a cubic array of
$N$ polarizable subunits, above a dielectric flat surface. The field
at each subunit can be written: \be
\label{dipi} \ve{E}(\ve{r}_i,\omega) & = &
\ve{E}_0(\ve{r}_i,\omega) + \sum_{j=1}^{N} [
\ve{S}(\ve{r}_i,\ve{r}_j,\omega)\nonumber\\ & + &
\ve{T}(\ve{r}_i,\ve{r}_j,\omega)] \alpha_j(\omega)
\ve{E}(\ve{r}_j,\omega).  \ee where $\ve{E}_0(\ve{r}_i,\omega)$ is the
field at the position $\ve{r}_i$ in the absence of the scattering
object, $\ve{T}$ and $\ve{S}$ are the field susceptibilities
associated to the free space~\cite{jackson} and the
surface,~\cite{agarwal,rahmani} respectively. $\alpha_i(\omega)$ is
the polarizability of the $i^{th}$ subunit. Like in
Ref.~[\ref{prbchaumetnieto}] we use the polarizability of the
Clausius-Mossotti relation with the radiative reaction term given by
Draine:~\cite{draine}
\be\label{cm}\alpha=\frac{\alpha_0}{1-(2/3)ik^3_0\alpha_0}\ee where
$\alpha_0$ holds the usual Clausius-Mossotti relation
$\alpha_0=a^3(\varepsilon-1)/(\varepsilon+2)$.~\cite{prbchaumetnieto,remarque}
In a recent paper,\cite{oplchaumetnieto} we have shown the importance
to compute the optical forces taking into account the radiative
reaction term in the equation for the polarizability of a sphere. For
a metallic sphere, the polarizability is written as
$\alpha=\alpha_0(1+(2/3)ik^3_0\alpha_0^*)/D$ with $D=1+(4/3)k_0^3\Im m
(\alpha_0)+(4/9)k_0^6|\alpha_0|^2$, where the asterisk stands for the
complex conjugate and $\Im m$ denotes the imaginary part.

The force~\cite{force} at each subunit is:~\cite{oplchaumetnieto}
\be\label{forcec} F_k(\ve{r}_i)=(1/2)\Re e\left(\alpha_i
{E_i}_l(\ve{r}_i,\omega) \left(\frac{\partial}{\partial k}
E^l(\ve{r},\omega)\right)_{\ve{r}=\ve{r}_i}^*\right), \ee where $k$
and $l$ stand for the components along either $x,y$ or $z$, and $\Re
e$ denotes the real part. The object is a set of $N$ small dipoles so
that it is possible to compute the force on each one from
Eq.~(\ref{forcec}). Hence, to obtain the total force on the particle,
it suffices to sum the contributions $\ve{F}(\ve{r}_i)$ on each
dipole.

\begin{figure}[H]  
\begin{center}
\resizebox{70mm}{!}{\input{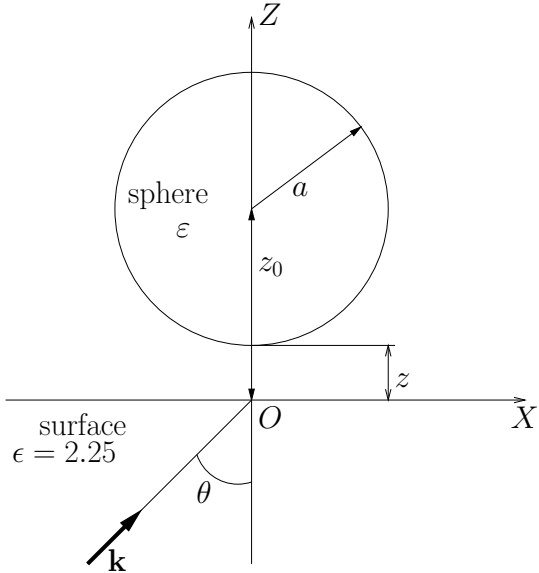}}
\end{center}
\caption{Geometry of the configuration considered. Sphere of radius
$a$ on a dielectric flat surface ($\epsilon=2.25$). The incident wave
vector $\ve{k}$ is in the $XZ$ plane.}
\end{figure}

Being the object under study a small sphere located at
$\ve{r}_0=(0,0,z_0)$ (see Fig.~1), we can employ the dipole
approximation, and hence use directly Eq.~(\ref{forcec}) with
$N$=1. Within the static approximation for the field susceptibility
associated to the surface (SAFSAS) (that is to say $k_0=0$), we have
found an analytical expression for $\ve{E}(\ve{r}_0,\omega)$ that
yields the force components:~\cite{prbchaumetnieto}
\be\label{force3dpa} F_x & = &\frac{\Re e}{2}\left[4\alpha z_0^3(i
k_x)^{*} \left(\frac{2|{E_0}_x|^{2}}{8z_0^{3}+\alpha\Delta}+
\frac{|{E_0}_z|^{2}}{4z_0^{3}+\alpha\Delta}\right)\right],\\
\label{force3dpb} F_z & = &
|{E_0}_x|^{2}\frac{\Re e}{2}\left( \frac{8z_0^3\alpha(i k_z)^{*}}
{8z_0^{3}+\alpha\Delta}+\frac{12z_0^2|\alpha|^2\Delta}
{|8z_0^{3}+\alpha\Delta|^{2}}\right)\nonumber\\& + &
|{E_0}_z|^{2}\frac{\Re e}{2}\left( \frac{4z_0^3\alpha(i k_z)^{*}}
{4z_0^{3}+\alpha\Delta}+\frac{6z_0^2|\alpha|^2\Delta}
{|4z_0^{3}+\alpha\Delta|^{2}}\right).\ee for $p$-polarization, and
\be\label{force3dsa} F_x & = & |{E_0}_y|^{2} \frac{\Re e}{2}\left[
\frac{8z_0^3\alpha(ik_x)^*}{8z_0^{3}+\alpha\Delta}\right],\\
\label{force3dsb} F_z & = & |{E_0}_y|^{2}\frac{\Re e}{2}
\left( \frac{8z_0^3\alpha(i k_z)^{*}}
{8z_0^{3}+\alpha\Delta}+\frac{12z_0^2|\alpha|^2\Delta}
{|8z_0^{3}+\alpha\Delta|^{2}}\right).\ee for $s$-polarization, with
$\Delta=(1-\epsilon)/(1+\epsilon)$ being the Fresnel coefficient of
the surface. We have assumed a dielectric surface, hence $\Delta$ is
real.

From Fig.~1 is easy to see that $k_x$ is always real whatever the
angle $\theta$, hence we can write Eqs.~(\ref{force3dpa})
and~(\ref{force3dsa}) for metallic particles as: \be\label{forcedippx}
F_x & = & \left(\frac{|{E_0}_x|^2 64z_0^6}{|8z_0^3+\alpha\Delta|^2}+
\frac{|{E_0}_z|^2 16z_0^6}{|4z_0^3+\alpha\Delta|^2}\right) \nonumber\\
& \times & [k_x\Im m (\alpha_0)/(2D)+k_x k_0^3|\alpha_0|^2/(3D)]\ee
for $p$ polarization, and \be\label{forcedipsx} F_x=\frac{|{E_0}_y|^2
64z_0^6}{|8z_0^3+\alpha\Delta|^2} [k_x\Im m (\alpha_0)/(2D)+k_x
k_0^3|\alpha_0|^2/(3D)]\ee for $s$-polarization. In
Eq.~(\ref{forcedippx}) the factor in front of $(k_x\Im m
(\alpha_0)/(2D)+k_x k_0^3|\alpha_0|^2/(3D))$ for the two polarizations
constitutes the field intensity at $z_0$. The first term within these
square brackets corresponds to the absorbing force whereas the second
represents the scattering force on the sphere. We see from
Eqs.~(\ref{forcedippx}) and~(\ref{forcedipsx}) that $F_x$ always has
the sign of $k_x$. Notice that it is not possible to write a general
equation for the force along the $Z$-direction, as $k_z$ will be
either real or imaginary, according to the angle of incidence.

\section{Results and discussion}\label{result}

\subsection{small particles}\label{petite}

We first address a small isolated silver particle with radius
$a=$10nm. In this case we can use the dipole approximation, hence we
consider Eqs.~(\ref{force3dpa})-(\ref{force3dsb}) with
$\Delta=0$. Fig.~2a presents the polarizability modulus ($|\alpha_0|$)
of the sphere. The maximum of the curve corresponds to the plasmon
resonance, i.e. when the dielectric constant is equal to -2 in Drude's
model. Notice that, in this model the dielectric constant is real, on
using experimental values,~\cite{palik} the dielectric constant is
complex and the resonance is not exactly at $\Re e(\varepsilon)=-2$
but slightly shifted. In Fig.~2b we plot the real part of the
polarizability ($\Re e(\alpha_0)$), and Fig.~2c shows its imaginary
part ($\Im m(\alpha_0)$). Fig.~2d represents the force in free space
computed from an exact Mie calculation (full line) and by the dipole
approximation from Eqs.~(\ref{force3dpa})-(\ref{force3dsb}) with
$\Delta=0$ (dashed line) and Eq.~(\ref{cm}) for $\alpha$. In this
case, the dipole approximation slightly departs from Mie's calculation
between 350nm and 375nm. We can compute the polarizability $\alpha$
from the first Mie coefficient $a_1$ given by Dungey and Bohren
(DB).~\cite{dungey} Therefore, the electric dipole polarizability is
$\alpha=3i a_1/(2k_0^3)$.~\cite{doyle} The symbol + in Fig.2d
corresponds to the DB polarizability and it is exactly coincident with
the Mie calculation. When the optical constant of the metallic sphere
is close to the plasmon resonance, the calculation from the
Clausius-Mossotti relation with the radiative reaction term, departs
from the exact calculation, even for a small radius. We shall next use
the polarizability of DB. Analytical calculations will always be done
with Eq.~(\ref{cm}) to get simple expressions, and thus a better
understanding of the physics involved. The curve of the force obtained
by Mie's calculation has exactly the same shape as the imaginary part
of the polarizability, this is due to the fact that for a small
metallic sphere the absorbing force is larger than the scattering
force.

\begin{figure}[H]  
\begin{center}
\includegraphics*[draft=false,width=80mm]{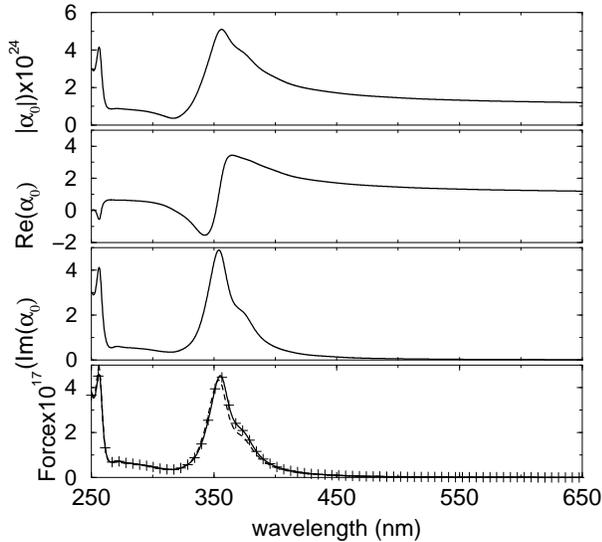}
\end{center}
\caption{ From top to bottom: the first three curves represent the
modulus, the real part, and the imaginary part of the polarizability
of a silver sphere with radius $a=10$nm versus the wavelength. The
fourth curve is the force on this particle in free space. Plain line:
Mie calculation, dashed line: polarizability of Clausius-Mossotti
relationm with the radiative reaction term, symbol +: DB
polarizability.}
\end{figure}

\begin{figure}[H]  
\begin{center}
\resizebox{80mm}{!}{\input{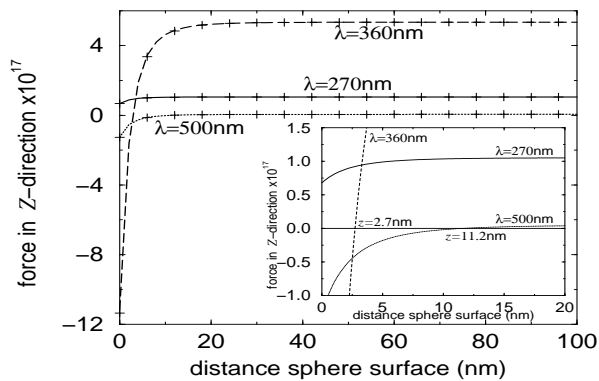}}
\end{center}
\caption{Force along the $Z$-direction on a silver sphere with
$a=10$nm versus distance $z$ in the dipole approximation. The angle of
incidence is $\theta=0^{\circ}$. With the static approximation (no
symbol) and in an exact calculation (+) for $\ve{S}$.  Dashed line
$\lambda=$360nm, plain line $\lambda=270$nm, and dotted line
$\lambda=$500nm. The inset shows details of the zero force.}
\end{figure}

Next, we consider the small sphere on a dielectric plane surface as
shown by Fig.~1. Illumination takes place from the dielectric side
with $\theta=0^{\circ}$, hence in vacuum $k_z=k_0$ and $k_x=0$. Fig.~3
represents the force in the $Z$-direction from Eq.~(\ref{force3dsb})
versus $z$ for different wavelengths. Far from the surface the force
tends to the Mie limit. Near the surface the force decreases, and,
depending on the wavelength, it can become negative. For a better
understanding of this force we write Eq.~(\ref{force3dsb}) as:
\be\label{force3dsb0}
F_z=\frac{64z_0^6|E_0|^2}{|8z_0^3+\alpha\Delta|^2}\left[\frac{k_0}{2}\Im
m(\alpha_0)+\frac{k_0^4}{3}|\alpha_0|^2+
\frac{3|\alpha_0|^2\Delta}{32z_0^4}\right], \ee having made the
approximation $|\alpha|\approx|\alpha_0|$, and $D\approx 1$ since we
have a small sphere compared to the wavelength ($k_0 a\ll 1$). The
factor in front of the bracket corresponds to the intensity of the
field at $z_0$. The first and the second terms within brackets
represent the interaction between the dipole moment associated to the
sphere and the incident field: the first term is the absorbing force
whereas the second one corresponds to the scattering force, hence
these forces are always positive. The third term is due to the
interaction between the dipole and the field radiated by the dipole
and reflected by the surface. We can consider this term as a gradient
force exerted on the sphere due to itself via the surface. Hence, this
force is always negative whatever the relative permittivity
$\varepsilon$. Since this term is proportional to $1/z_0^4$, it
becomes more dominant as the sphere approaches the surface, hence the
force decreases. To derive the point $z_0$ at which the force
vanishes, if such a point exists, let us assume the scattering force
smaller than the absorbing force, then from Eq.~(\ref{force3dsb0}) the
zero force is: \be
\label{zeroforce} z_0^4=\frac{3|\alpha_0|^2}{16 k_0 \Im
m(\alpha_0)}\frac{\epsilon-1}{\epsilon+2}.\ee This equation always has
a solution. We find $z_0$ for the three wavelengths used to be:
$\lambda=270$nm, $z_0=7.9$nm, $\lambda=360$nm, $z_0=12.8$nm,
$\lambda=500$nm, $z_0=21.6$nm. Now, $z_0$ (the location of the center
of the sphere) must be larger than the radius $a$, or else, the sphere
would be buried in the surface. Therefore the first of those values of
$z_0$ is not possible. Hence, the force is always positive. Thus, the
distance between the sphere and the surface is $z$=2.8nm and 11.6nm
for $\lambda=$360nm and 500nm, respectively. These values are very
close to those shown in the inset of Fig.~2. Notice that near the
plasmon resonance for the sphere ($\lambda \approx 360$nm) both
$|\alpha_0|$ and $\Im m(\alpha_0)$ are maxima, hence the force is very
large when the sphere is far from the surface and, due to the third
term of Eq.~(\ref{force3dsb0}), which depends of $|\alpha_0|^2$, the
decay of this force is very fast. We have used the static
approximation for the susceptibility tensor associated to the surface
(SAFSAS) at large distance. We plot in Fig.~2 with crosses the force
obtained from an exact calculation for $\ve{S}$ with the dipole
approximation. As we see, these crosses coincide with those curves
obtained with the SAFSAS whatever the distance $z$. Yet, we obtain for
dielectric spheres the following: near the surface the SAFSAS is
valid, whereas far from the surface the sphere does not feel its
presence, and, thus, whether using the tensor susceptibility
associated to the surface in its exact form, or within the static
approximation, has no influence.

\begin{figure}[H]  
\begin{center}
\includegraphics*[draft=false,width=80mm]{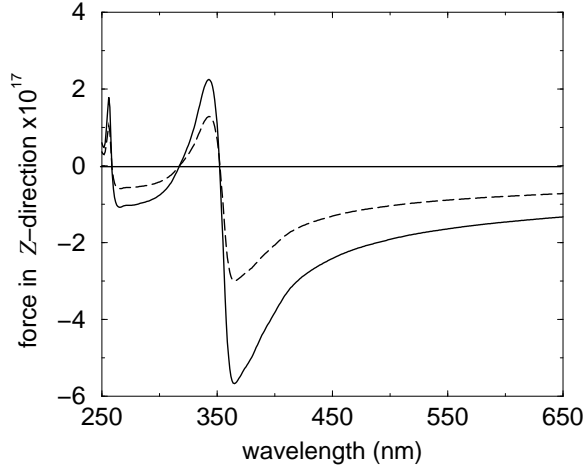}
\end{center}
\caption{Force along the $Z$-direction on a silver sphere with
$a=10$nm versus the wavelenght $\lambda$ in the dipole
approximation. The angle of incidence is $\theta=50^{\circ}$. Plain
line: $p$-polarization; dashed line: $s$-polarization.}
\end{figure}

\begin{figure}[H]  
\begin{center}
\includegraphics*[draft=false,width=80mm]{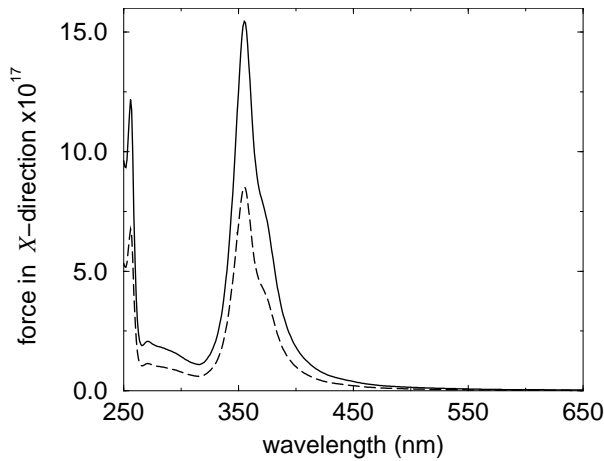}
\end{center}
\caption{Force along the $X$-direction on a silver sphere with
$a=10$nm versus the wavelenght $\lambda$ in the dipole
approximation. The angle of incidence is $\theta=50^{\circ}$. Plain
line: $p$-polarization; dashed line: $s$-polarization.}
\end{figure}

We next consider the surface illuminated at angle of incidence
$\theta$ larger than the critical angle:
$\theta=50^{\circ}>41.8^{\circ}=\theta_c$. Now the transmitted
electromagnetic wave above the surface is evanescent. We plot in
Fig.~4 from Eqs.~(\ref{force3dpa})-(\ref{force3dsb}) the force on the
sphere for the two polarizations in the $Z$-direction versus the
wavelength, and in the $X$-direction in Fig.~5, when the sphere is
located at $z_0$=30nm. In Fig.~4 we see that the force in the
$Z$-direction is also either positive or negative. In a previous
work~\cite{prbchaumetnieto}, we have observed that the force on a
small dielectric sphere is always attractive when the sphere is
located in an evanescent wave. This is no longer the case for a
metallic sphere. To understand this difference, and as the two
polarizations have the same behaviour, we take the analytical solution
for $F_z$ with $k_z=i\gamma$ ($\gamma>0$) for $s$-polarization. Then
Eq.~(\ref{force3dsb}) can be written: \be F_z & = &
\frac{|{E_0}_y|^2}{|8z_0^3+\alpha\Delta|^2}\frac{\Re e}{2} \Big(
-\gamma 8 z_0^3\alpha(8z_0^3+\alpha^*\Delta)\nonumber\\ & + &
12z_0^2|\alpha|^2\Delta\Big)\ee On using the approximation $D\approx
1$ and $|\alpha|\approx|\alpha_0|$ we obtain: \be \label{forceevas}
F_z=\frac{64 z_0^6|{E_0}_y|^2}{|8z_0^3+\alpha\Delta|^2}
\left(-\frac{\gamma\Re
e(\alpha_0)}{2}-\frac{\gamma|\alpha_0|^2\Delta}{16z_0^3}+
\frac{3|\alpha_0|^2\Delta}{32z_0^4}\right)\ee As shown by Fig.~3, when
the sphere is located at $z_0=$30nm, the influence of the surface
becomes negligible, thus we can use Eq.~(\ref{forceevas}) with the
hypothesis that $z_0$ is large. Hence, $F_z\approx -|{E_0}_y|^2 \gamma
\Re e(\alpha_0)/2$. This is the gradient force due to the incident
field, and therefore due to the interaction between the dipole
associated to the sphere and the applied field. This force exactly
follows the behaviour of $\Re e(\alpha_0)$ (cf. Fig.~2).  When $\Re
e(\alpha_0)$ is negative, the dipole moment of the sphere oscillates
in opposition to the applied field and so the sphere is attracted
towards the weaker field. Notice that the same phenomenon is used to
build an atomic mirror: for frequencies of oscillation higher than the
atomic frequency of resonance, the induced dipole oscillates in phase
opposition with respect to the field.  The atom then undergoes a force
directed towards the region of weaker field.~\cite{aspect} For
$p$-polarization, the force can be written $F_z\approx
-(|{E_0}_x|^2+|{E_0}_z|^2) \gamma \Re e(\alpha_0)/2$. As the modulus
of the field becomes more predominant in $p$-polarization, the
magnitude of the force becomes more important. We now search more
carefully the change of sign in the force. Writing
$\varepsilon=\varepsilon'+i\varepsilon''$ for the relative
permittivity, we get: \be\label{reelpartie} \Re e(\alpha_0) & = & a^3
\frac{(\varepsilon'-1)(\varepsilon'+2)+\varepsilon''^2}
{(\varepsilon'+2)^2+\varepsilon''^2}\\ \label{imapartie} \Im
m(\alpha_0) & = & a^3
\frac{3\varepsilon''}{(\varepsilon'+2)^2+\varepsilon''^2} \ee If the
damping is weak, then the change of sign of $F_z$ happens both for
$\varepsilon'\approx 1$ and at the plasmon resonance for the sphere,
i.e., $\varepsilon'\approx -2$. Between these two values, the gradient
force is positive. In fact, the limiting values of the positive
gradient force are always strictly in the interval $[-2, 1]$ due to
damping. For example, the force vanishes at $\lambda=352$nm with
$\varepsilon=-1.91+0.6i$ and $\lambda=317$nm with
$\varepsilon=0.66+0.95i$. We notice that the change of sign happens
steeply at the plasmon resonance since then the denominator of the
real part of the polarizability becomes very weak (see
Eq.~(\ref{reelpartie})), hence the zero force is surrounded by the two
maxima of the force (one positive and the other negative). At
$\lambda=317$nm the change of sign is smoother as in that case the
denominator is far from zero. The third case, $\lambda=259$nm, lies
between those two previous cases as the damping of the relative
permittivity is important: $\varepsilon=-1.65+1.12i$. We have also
investigated the cases of gold and copper spheres, where a plasmon
easily takes place, but we found not possible change of sign for $F_z$
for these two materials as the damping is then too important: if
$\varepsilon''>3/2$, then $\Re e(\alpha_0)$ is always positive
whatever $\varepsilon'$. However, if the particle is embedded in a
liquid with a relative permittivity 2, then it is possible to get $\Re
e(\alpha)<0$ for gold. Notice that if $\theta$ is close to $\theta_c$,
then $\gamma\approx 0$ and so we only have the third term of
Eq.~(\ref{forceevas}), then the force is always negative whatever the
wavelength. In Fig.~5 we see, as previously, that the force in the
$X$-direction has the sign of $k_x$ and, as the absorbing force is the
most predominant one, the curve has the same shape as the imaginary
part of the polarizability (cf. Fig.~2). In that case, the maximum of
the force $F_x$ is at the plasmon resonance (see
Eq.~(\ref{imapartie})) namely at $\lambda=354$nm.

\begin{figure}[H]  
\begin{center}
\includegraphics*[draft=false,width=80mm]{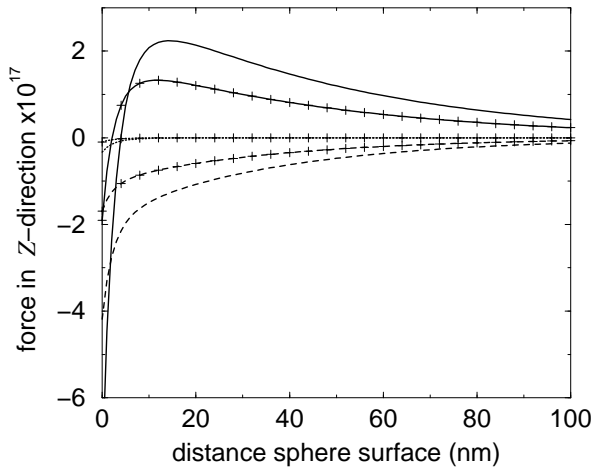}
\end{center}
\caption{Force along the $Z$-direction on a silver sphere with
$a=10$nm versus distance $z$ in the dipole approximation. The angle of
incidence is $\theta=50^{\circ}$. Without symbol: $p$-polarization;
with cross (+): $s$-polarization. Plain line: $\lambda=$340nm, dashed
line: $\lambda$260=nm, dotted line: $\lambda=$317.5nm.}
\end{figure}

Fig.~6 shows the force in the $Z$-direction versus $z$ for both $s$
(symbol +) and $p$-polarization (without any symbol) at three
different wavelengths ($\lambda=$265, 340, 317.5nm) for
$\theta=50^{\circ}$. The behaviour is the same for both polarizations,
only appearing a difference of magnitude, this is due to the component
of the field perpendicular to the surface in $p$-polarization. All
curves manifest that near the surface the force is attractive, this is
due, as seen before with a propagating wave, to the term of
$\Delta/z_0^4$ in Eqs.~(\ref{force3dsb0}) and (\ref{forceevas}) which
is always negative, irrespective of the kind of wave above the
surface. At $\lambda=$317.5nm ,(dotted line), we have $\Re
e(\alpha)=0$ which is why the force very quickly goes to zero when z
grows. The two other cases correspond to $\Re e(\alpha)>0$
($\lambda=$260nm) and $\Re e(\alpha)<0$ ($\lambda=$340nm) and far from
the surface the force tends to zero. As the force is proportional to
$|\ve{E}_0|^2$, we have $F_z\propto e^{(-2\gamma z)}$.

\subsection{large particles}\label{grande}

It is difficult to obtain convergence for the CDM when the relative
permittivity of the medium to discretize is large. This imposes a very
fine sampling. In this section, we use the range 250-355nm for the
wavelength, the real part of the relative permittivity being
small. Then, the computation is simplified. In addition, this is the
most interesting case as then $\Re e(\alpha)=0$ three times in this
interval of wavelengths.

\begin{figure}[H]  
\begin{center}
\includegraphics*[draft=false,width=80mm]{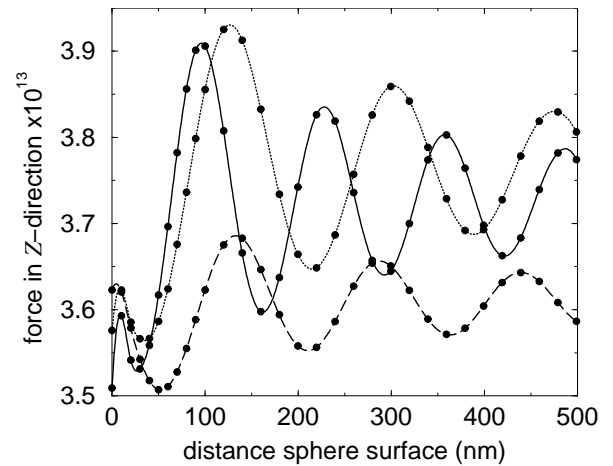}
\end{center}
\caption{Force along the $Z$-direction on a silver sphere versus
distance $z$ with $\theta=0$ and $a$=100nm for the following
wavelengths: $\lambda=255$nm (plain line), $\lambda=300$nm (dashed
line), and $\lambda=340$nm (dotted line). Dots correspond to computed
points.}
\end{figure}

In Fig.~7 we plot the force in the $Z$-direction for an incident
propagating wave ($\theta=0^{\circ}$) versus the distance between the
sphere and the surface at three different wavelengths:
$\lambda=$255nm, 300nm, 340nm. The calculations are done without any
approximation. The curves have a similar magnitude and behaviour at
the three wavelengths. The forces present oscillations due to the
multiple reflection of the radiative waves between the sphere and the
surface, hence the period of these oscillations is $\lambda/2$. The
magnitude of these oscillations depends on the reflectivity of the
sphere, so the higher the Fresnel coefficient is, the longer these
oscillations are. As expected, they are less remarkable when the
sphere goes far from the surface. We notice that the decay of the
force when the metallic sphere gets close to the surface is not
comparable to that on a dielectric sphere (see
Ref.[\ref{prbchaumetnieto}]). This is due to strong absorbing and
scattering forces on the metallic sphere in comparison to the gradient
force induced by the presence of the dielectric plane.

\begin{figure}[H]  
\begin{center}
\includegraphics*[draft=false,width=80mm]{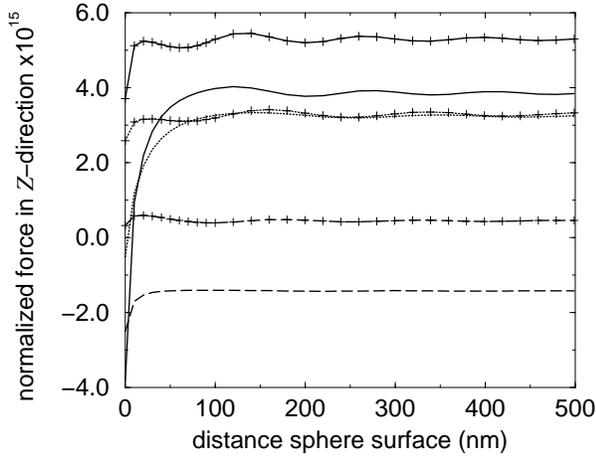}
\end{center}
\caption{Force along the $Z$-direction on a silver sphere with
$a$=100nm versus distance $z$ with $\theta=50^{\circ}$ for the
following wavelengths: $\lambda=255$nm (plain line), $\lambda=300$nm
(dashed line), and $\lambda=340$nm (dotted line). symbol +:
$s$-polarization; without symbol: $p$-polarization.}
\end{figure}

In Fig.~8 we plot for $\theta=50^{\circ}$ the normalized force in the
$Z$-direction, i.e., $F_z/|\ve{E}_0|^2$, $\ve{E}_0$ being the field at
$z_0$ in the absence of the sphere. We remark two important facts at
this angle of incidence. First, the decay of the force when the sphere
is near the surface is more important in $p$-polarization. Notice that
with the CDM it is not possible to numerically split the scattering,
absorbing, and gradient forces. Therefore, when the sphere is large we
shall argument on the set of dipoles forming it. In $p$-polarization,
due to the $z$-component of the incident field, the dipoles also have
a component perpendicular to the surface, and this is larger than in
$s$-polarization.  Hence, in agreement with Fig.~6, due to this
$z$-component, the attraction of the sphere towards the surface is
larger for $p$-polarization. Second, all forces are positive when the
sphere is far from the surface except for $\lambda=$300nm in
$p$-polarization. At this wavelength, for a small sphere, the force is
negative for both $s$ and $p$-polarization, hence we can assume an
effect due to the size of the sphere. Just to see this effect, we
present in Fig.~9 the force in the $Z$-direction versus the radius
$a$, on a sphere located at $z_0=100$nm for an angle of incidence
$\theta=50^{\circ}$ but without taking into account the multiple
interaction with the surface (i.e., $\ve{S}=0$). We take the previous
wavelength of Fig.~8 ($\lambda=255$nm, 300nm, and 340nm) more the
wavelength at the plasmon resonance, $\lambda=351.5$nm where $\Re
(\alpha_0)=0$.  For small radius, we observe the same behaviour as in
the previous section for an incident evanescent wave, namely at
$\lambda=255$nm and 340nm, the force in the $Z$-direction is positive,
and it becomes negative for $\lambda=300$nm for both
polarizations. These curves show a dependence proportional to the cube
of the radius as $\Re (\alpha_0)\propto a^3$. At the plasmon
resonance, the force is slightly positive as there is no gradient
force, but only weak absorbing and scattering forces. 

\begin{figure}[H]  
\begin{center}
\includegraphics*[draft=false,width=80mm]{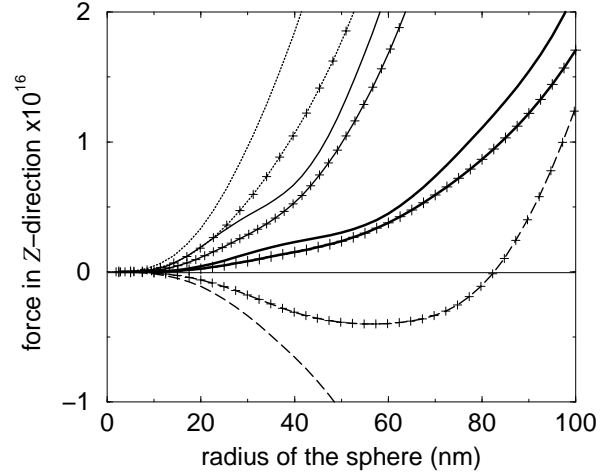}
\end{center}
\caption{Force along the $Z$-direction on a silver sphere located at
$z_0=$100nm, with $\theta=50^{\circ}$, versus the radius $a$ for
$\lambda=255$nm (plain line), $\lambda=300$nm (dashed line),
$\lambda=340$nm (dotted line), and $\lambda=351.5$nm (thick
line). Symbol +: $s$-polarization; without symbol:
$p$-polarization. The interaction between the sphere and the surface
is not taken into account.}
\end{figure}

But as shown by Fig.~9, when the radius grows, in $s$-polarization at
$\lambda=$300nm, the force sign changes and it becomes positive around
$82$nm as $p$-polarization keep the same behaviour. This confirms the
fact that the positive force obtained in Fig.~8 for $\lambda=$300nm is
only a size effect.  For the cases $\lambda=255$nm, $340$nm, the
gradient force is positive in the $Z$-direction, like the scattering
and absorbing forces . Hence, the force, is always positive whatever
the radius. Nevertheless, for $\lambda=300$nm, these is a negative
gradient force, the two other forces being positive. As previously
said, it is not possible to know the relative contribution of the
different forces, but in $s$-polarization the dipoles have a component
mainly parallel to the plane. In that case, the dipole associated to
each subunit radiates a field both propagating and evanescent along
the $Z$-direction. Hence, we can assume that due to the radiative
part, the absorbing and scattering forces acting on the subunits
inside the sphere, become important when the radius
increases. Concerning $p$-polarization, due to the component of the
field perpendicular to the plane, the dipoles have a component along
$Z$ and thus only radiate evanescent waves along this direction. In
this case, we believe that the gradient force is more important on
each subunit and counterbalances the absorbing and scattering
forces. At plasmon resonance $\lambda=351.5$nm, when the radius grows,
the absorbing and scattering forces become larger, but as no gradient
force exists, the force is lower than those obtained at
$\lambda=255$nm and 340nm. In fact, the curve shown at $\lambda=340$nm
is the most important one, due to the onset of the plasmon resonance,
thus $|\alpha_0|$ and $\Im m(\alpha_0)$ being near their maximum, and
$\Re e(\alpha_0)$ close to its minimum. In this case, the gradient
force is maximum and positive.

\section{Conclusions}\label{conclu}

We have presented a theoretical study of the optical forces acting
upon a metallic particle on a dielectric plane surface either
illuminated at normal incidence or under total internal
reflection. This study is done both with the coupled dipole method and
Maxwell's stress tensor. We observe that when the incident wave is
propagating, the difference between the force acting on a dielectric
sphere and that on a metallic sphere stems from the absorbing
force. Due to this contribution, the force upon a small silver sphere
close to the dielectric surface can be positive in spite of the
gradient force.  The opposite happens with a dielectric sphere. The
main difference between the two cases (dielectric and metallic) arises
however on illumination under total internal reflection. In that case,
the effect on a small silver sphere is completely different to that
observed on a dielectric sphere. Depending on the wavelength, the
gradient force due to the incident field can be either repulsive or
attractive. The change of sign happens both at the plasmon resonance
and when $\varepsilon$ becomes close to one. In the interval between
these two values the gradient force is positive. The explanation is
very similar to that on the effect used to build an atomic mirror. At
a wavelength where the gradient force on a small sphere is negative,
we see that when the sphere radius grows, the force along the
$Z$-direction stays negative for $p$-polarization, and it becomes
positive for $s$-polarization due to the size effect. Nevertheless, at
any arbitrary wavelength and angle of incidence, as the sphere
approaches close to the surface, the attraction of the surface on the
sphere increases: repulsive forces diminish and even can change sign,
eventually becoming attractive at certain wavelengths. Attractive
forces, on the other hand, increase their magnitude.

\section{Acknowledgments}

Work supported by the European Union and Direcci\'on General de
Investigaci\'on Cientif\'{\i}ca y T\'ecnica. P. C. acknowledges
financial support from a TMR contract of the European Union.


\end{multicols}


\begin{references}
\bibitem{remarque} $a^3=3d^3/(4\pi)$ where $d$ is the spacing of
lattice discretization. In the case of the dipole approximation $a$ is
the radius of the sphere.
\bibitem{force} In fact $\ve{F}$ is not exactly the force but moreever
the time averaged force.
\end{references}
\end{document}